\documentclass[12pt,preprint]{emulateapj}
\usepackage{apjfonts}
\usepackage[dvipdfm,colorlinks,linkcolor=red,anchorcolor=black,citecolor=blue]{hyperref}
\begin{document}

\title{From outside-in to inside-out: galaxy assembly mode depends on stellar mass}
\shortauthors{Pan et al.}
\shorttitle{galaxy assembly mode depends on mass}
\author{Zhizheng Pan\altaffilmark{1}, Jinrong Li\altaffilmark{2,3}, Weipeng Lin \altaffilmark{1,5}, Jing Wang\altaffilmark{4}, Lulu Fan\altaffilmark{2,3}, Xu Kong\altaffilmark{2,3} }

\email{panzz@shao.ac.cn, linwp@shao.ac.cn,xkong@ustc.edu.cn}

\altaffiltext{1}{Key laboratory for research in galaxies and cosmology, Shanghai Astronomical Observatory,
Chinese Academy of Science, 80 Nandan Road, Shanghai, 200030, China}

\altaffiltext{2}{Center of Astrophysics, University of
Science and Technology of China, Jinzhai Road 96, Hefei 230026, China}
\altaffiltext{3}{Key Laboratory for Research in Galaxies and Cosmology, USTC,
CAS, China}
\altaffiltext{4}{CSIRO Astronomy \& Space Science, Australia Telescope National Facility, PO Box 76, Epping, NSW 1710, Australia}
\altaffiltext{5}{School of Astronomy and Space Science, Sun Yat-Sen University, Guangzhou, 510275, China}

\begin{abstract}
In this Letter, we investigate how galaxy mass assembly mode depends on stellar mass $M_{\ast}$, using a large sample of $\sim$10, 000 low redshift galaxies. Our galaxy sample is selected to have SDSS $R_{90}>5\arcsec.0$, which allows the measures of both the integrated and the central NUV$-r$ color indices. We find that: in the $M_{\ast}-($ NUV$-r$) green valley, the $M_{\ast}<10^{10}~M_{\sun}$ galaxies mostly have positive or flat color gradients, while most of the $M_{\ast}>10^{10.5}~M_{\sun}$ galaxies have negative color gradients. When their central $D_{n}4000$ index values exceed 1.6, the $M_{\ast}<10^{10.0}~M_{\sun}$ galaxies have moved to the UV red sequence, whereas a large fraction of the $M_{\ast}>10^{10.5}~M_{\sun}$ galaxies still lie on the UV blue cloud or the green valley region. We conclude that the main galaxy assembly mode is transiting from "the outside-in" mode to "the inside-out" mode at $M_{\ast}< 10^{10}~M_{\sun}$ and at $M_{\ast}> 10^{10.5}~M_{\sun}$. We argue that the physical origin of this is the compromise between the internal and the external process that driving the star formation quenching in galaxies. These results can be checked with the upcoming large data produced by the on-going IFS survey projects, such as CALIFA, MaNGA and SAMI in the near future.

\end{abstract}

\section{Introduction}
The stellar mass of a galaxy, $M_{\ast}$, is one of its most fundamental properties. The assembly of $M_{\ast}$ is a long-standing issue in the field of galaxy formation and evolution. To elucidate how $M_{\ast}$ evolves, one would expect to recover the star formation history (SFH), both in space and time, for individual galaxies. In practice, the SFH of a galaxy is determined by finding the most plausible combination of evolved single stellar populations (SSP) that matches its observed spectrum or spectral energy distribution (SED). This so-called "fossil record method" has been widely applied to the integral field spectroscopic (IFS) data and the multi-band images with good spatial coverage to study galaxy assembly \citep{Kong 2000, Lin 2013, perez 2013, Cid 2013}.

Previous works have found two distinct stellar mass assembly modes. They are termed as "the inside-out" and "the outside-in" modes in the literature \citep{Sanchez 2007, perez 2013}. In the "inside-out" scenario, mass assembly is firstly finished in the galactic central region. For instance, in a young disk galaxy, the disk instability will induce gas inflow and trigger starburst in the galactic center as it evolves. Besides of star formation, the inflow gas can also fuel the central supermassive blackhole. The subsequent AGN feedback (such as blowing the gas out of the galaxy, or heating it against cooling down to form new stars) will then suppress the central star formation, leaving a compact, quiescent galactic bulge \citep{Dekel 2014,Bournaud 2014}. Compared to the central part, its outskirts form through much gentler process, such as the star formation driven by the gradually accretion of cold inter-galactic medium (IGM), or the accretion of small satellite galaxies. Many works found that massive spiral galaxies have negative color or stellar age gradients, supporting this picture \citep{Wang 2011, perez 2013,Lin 2013, Sanchez 2014}. Compared to their massive counterparts, low mass galaxies have much shallower potential wells, making them less capable to retrieve the gas blowing by the galactic outflows, or accrete cold gas to form new stars. In addition, the role of AGN feedback may not be so efficient in low mass galaxies, as their AGN occupation fraction is very low \citep{Kauffmann 2003b}. Thus it is not surprised to see the low mass galaxies have a different mass assembly mode. In fact, observations show that the evolution of low mass galaxies can be better interpreted in an "outside-in" framework.  \citep{Gallart 2008, Zhang 2012}

It is now clear that the galaxy assembly mode has dependence on its already formed $M_{\ast}$. Nevertheless, we know less about whether there is/or not a mass threshold which can roughly separate these two distinct modes. Answering this question requires the investigation of a sufficient large sample, which expanding a wide mass range. Recently, using the IFS data of 105 galaxies observed by the CALIFA project \citep{Sanchez 2012}, \citet{perez 2013} found that galaxies that are more massive than 5$\times10^{9.0}M_{\odot}$ grow inside$-$out,  while lower mass galaxies grow outside$-$in. However, this sample is still relatively modest, and there are only about 20 galaxies with $M_{\ast}<10^{10}M_{\odot}$ galaxies in this sample (see their Figure.1). To study a larger sample, in this Letter, we use the \emph{GALEX} \citep{Martin 2005} photometry, combined with the SDSS \citep{York 2000} data, to investigate how galaxy assembly mode depends on stellar mass.  Throughout this Letter, we assume a concordance $\Lambda$CDM cosmology with $\Omega_{\rm m}=0.3$, $\Omega_{\rm \Lambda}=0.7$, $H_{\rm 0}=70$ $\rm km~s^{-1}$ Mpc$^{-1}$, and a \citet{Kroupa 2001} IMF.

\section{Method and data used}
In this Letter, we will investigate the stellar age in the central region, as well that of the global, in individual galaxies. For a galaxy, by comparing these two quantities, we can investigate whether its central part is first assembled. Traditionally, stellar age is determined using broad-band optical photometry when spectral information is not available. However, the use of optical photometry has been plagued by the well$-$known age$-$metallicity degeneracy (AMD) whereby young, metal-rich stellar populations produce optical colors which are indistinguishable from those produced by
old, metal-poor populations \citep{Worthey 1994}. To avoid the AMD, We use the NUV$-r$ color index as an stellar age indicator, since it is sensitive to low level recent star formation (RSF), but is unaffect by the AMD when stellar age is < 1 $Gyr$ \citep{Kaviraj 2007a, Kaviraj 2007b}.

Following \citet{Wang 2010}'s pipeline, we have constructed a UV-optical matched photometric catalog. This catalog contains about 220,000 galaxies with uniform photometric measurements on the resolution and PSF-matched \emph{GALEX}+SDSS images, which are cross-matched by the SDSS DR8 \citep{Aih11} spectroscopic galaxies and the \emph{GALEX} GR6 database.  For each galaxy, the fluxes were measured over 5 different apertures, with $r=$[1.5, 3.0, 6.0, 9.0, 12.0] \arcsec, in the $FUV, NUV, u, g, r, i, z$ bands. Along with the aperture photometry, we also measure the total magnitudes of the galaxies with \textsc{SExtractor} \citep{Bertin 1996}. All the magnitudes have been corrected for galactic extinction using the galactic dust map \citep{Schlegel 1998}. The data reduction procedure is refereed to \citet{Wang 2010} and \citet{Pan 2014}.

To measure the NUV$-r$ color index in the central region of a galaxy, one would expect the galaxy to be spatially resolved in both the \emph{GALEX} and SDSS images. The \emph{GALEX} NUV image has a pixel size of 1 pixel=$1\arcsec. 5$ and a point spread function (PSF) with full width at half-maximum (FWHM)=$5\arcsec. 3$. The pixel size and PSF of the SDSS image are $0\arcsec .396$ and $1\arcsec .4$, respectively. To ensure that the measured central NUV$-r$ color index is not significantly affected by the relative poor resolution of the \emph{GALEX} images, the chosen central aperture should not be smaller than the NUV PSF, e.g, $r_{\rm central}$ should not be smaller than $5\arcsec. 3/2\approx$ 2\arcsec.7.  In the 5 apertures, the $r=3\arcsec.0 $ aperture is thus most suitable to be chosen as the "central aperture". Then, we select the galaxies passing the following as our parent sample: 1) SDSS minor$-$major axis ration $b/a>0.5$ ;2) SDSS $R_{90}>5.0 \arcsec$, where $R_{90}$ is the radii enclosed 90\% of the SDSS r-band petrosian flux; 3) $z=[0.005, 0.05]$, where $z$ is SDSS spectroscopic redshift; 4)$m_{\rm NUV}<m_{\rm limit}$, where $m_{\rm limit}=23.0~$mag is the limited magnitude in the NUV band. The limit on $b/a$ is to minimize the dust reddening effect on our results. Finally, we limit these galaxies to have stellar mass $M_{\ast}>10^{9.0}~M_\sun$, to form a volume-completed sample at $z<0.05$ \citep{Schawinski 2010}. The final sample size is $N_{\rm gal}$=11, 294. The stellar mass $M_{\ast}$, and the spectral index $D_{n}4000$ used in this Letter was drawn from the JHU/MPA database \footnote{http://www.mpa-garching.mpg.de/SDSS/DR7}.

In the upper pannel of Figure.1, we show the central flux fraction ($f_{\rm central}$) distribution for our sample, where $f_{\rm central}$ was defined as $f_{\rm central}=Flux_{\rm r=3\arcsec.0}/Flux_{\rm total}$.  It can be seen that in the SDSS r-band, the median value of $f_{\rm central}$ is $\sim$ 0.25. One will also find that the $f_{\rm central}$ distribution is much broader in the NUV band. The Kolmogorov-Smirnoff (KS) test on these two distributions gives a small probability (<0.01) that they are drawn from the same distribution. This may be due to the highly different RSF in the galaxy central regions. Figure.1 illustrates that, even for the $R_{90}\approx 5.0 \arcsec$ galaxies, in the SDSS r-band, their $f_{\rm central}$ are not higher than 0.4. Thus for this galaxy sample, our chosen central aperture should be appropriate.


\begin{figure}
\centering
\includegraphics[width=80mm,angle=0]{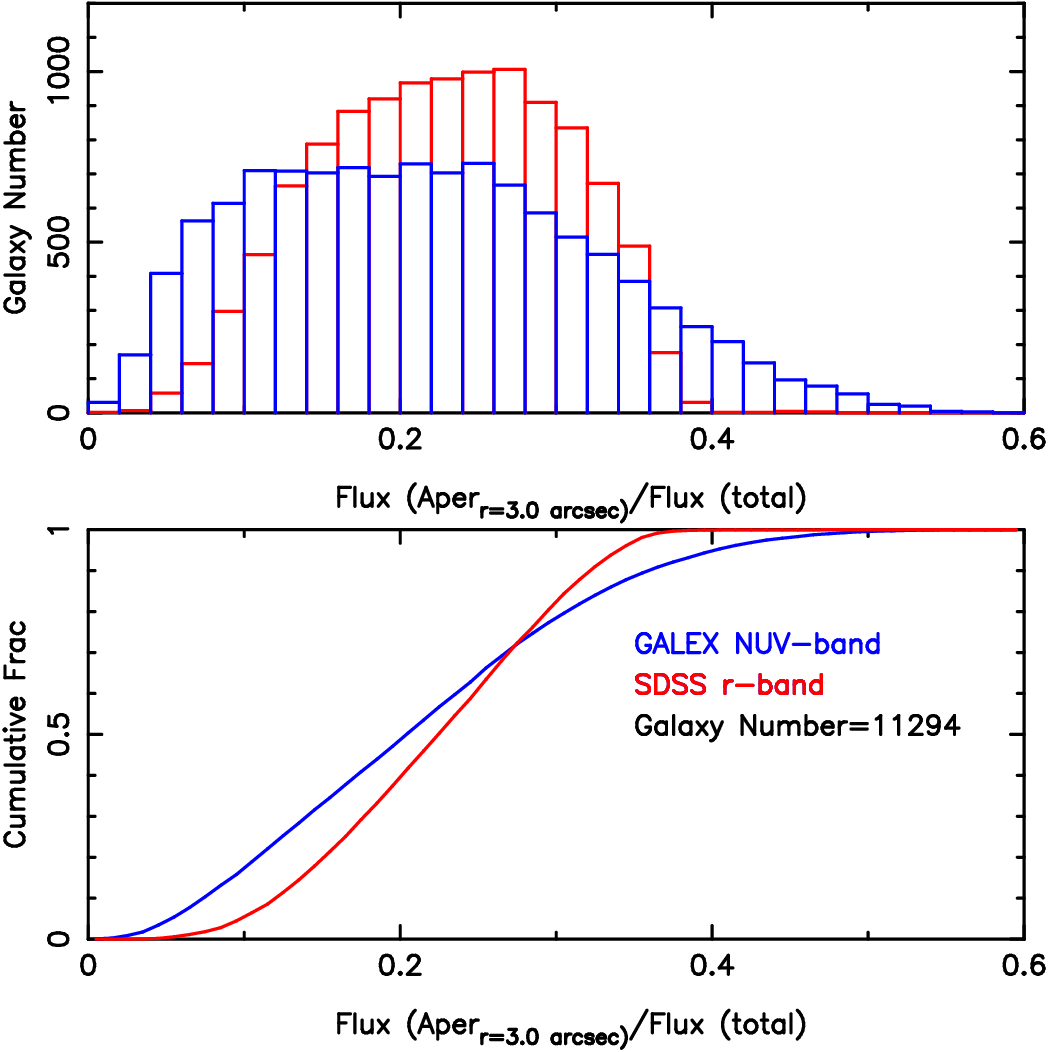}
\caption{Upper panel: The $f_{\rm central}$ distribution for our sample, where $f_{\rm central}$ was defined as $f_{\rm central}=Flux_{\rm r=3\arcsec.0}/Flux_{\rm total}$. The \emph{GALEX} NUV and SDSS r band is shown in blue and red histograms, respectively. Lower panel: the cumulative curve of $f_{\rm central}$. }
\end{figure}

\begin{figure*}
\centering
\includegraphics[width=120mm,angle=0]{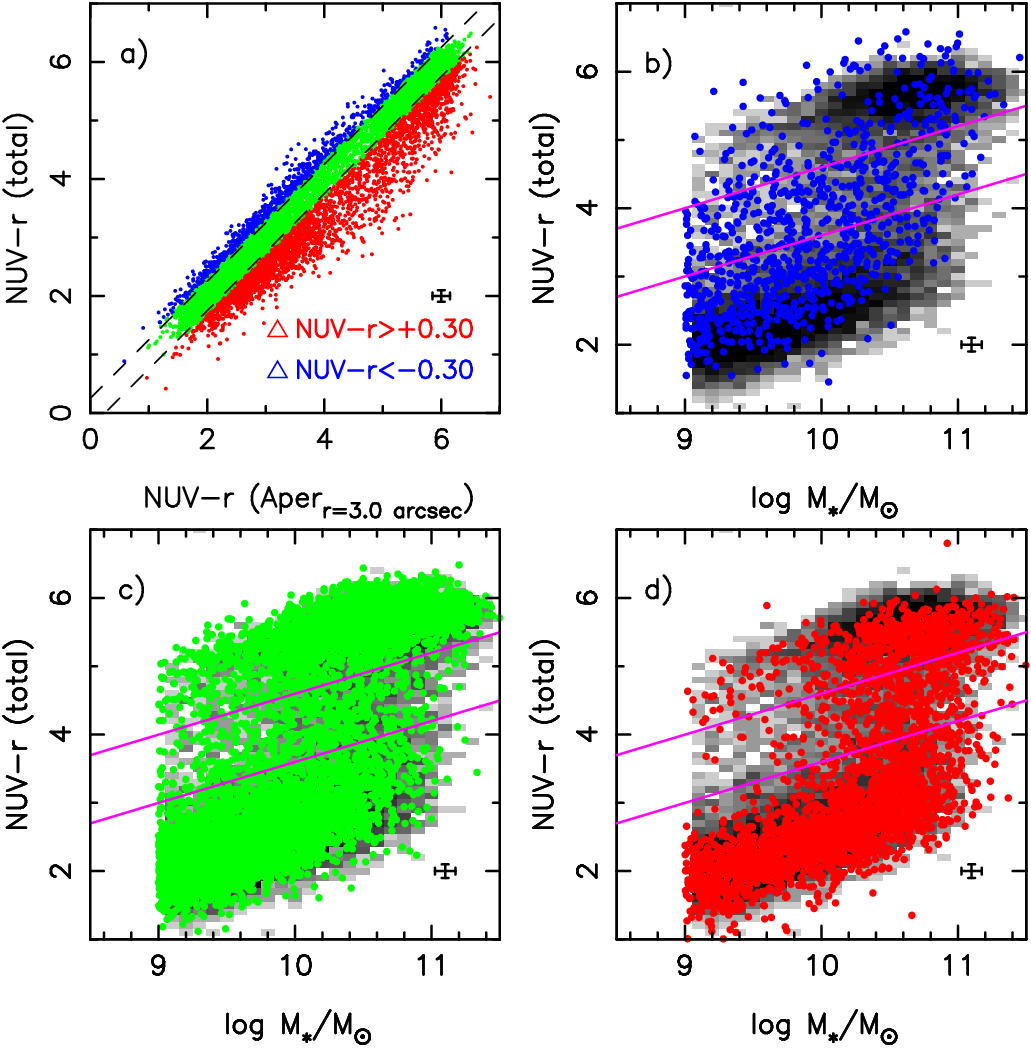}
\caption{a):The relation between NUV$-r_{\rm r=3\arcsec .0}$ and NUV$-r_{\rm total}$, where NUV$-r_{\rm r=3\arcsec .0}$ and NUV$-r_{\rm total}$ are the color indices measured in the central $r=3\arcsec .0$ aperture and that of the whole galaxy, respectively.  $\Delta$NUV$-r$ is defined as $\Delta$NUV$-r=$  (NUV$-r_{\rm r=3\arcsec .0}$)$-$(NUV$-r_{\rm total}$).  b): The $M_{\ast}$ .vs NUV$-r$ (total) diagram for galaxies with $\Delta$NUV$-r <-0.3$. c): CMD for galaxies with  $-0.3<\Delta$ NUV$-r<0.3$; d): CMD for galaxies with $\Delta$ NUV$-r>0.3$. The two pink lines denote our "green valley" definition.}
\end{figure*}

\begin{figure*}
\centering
\includegraphics[width=130mm,angle=0]{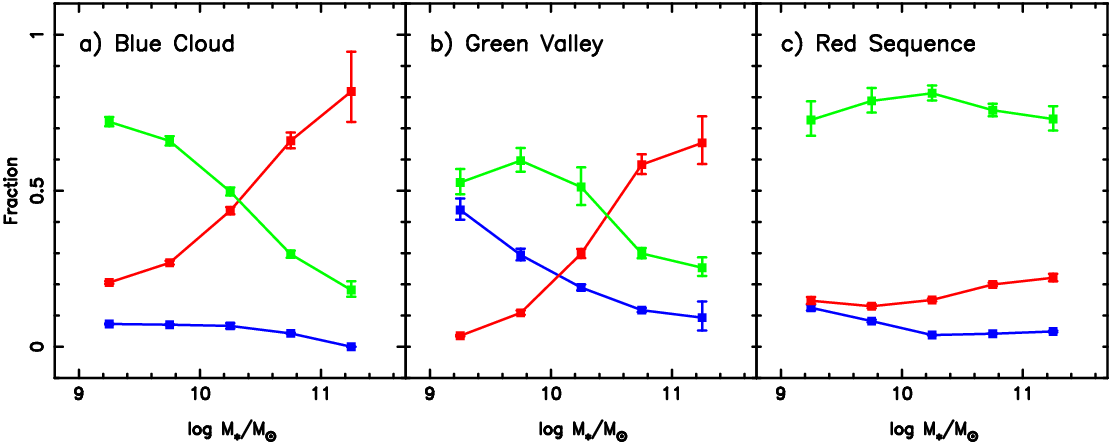}
\caption{The fractions of blue-cored (blue symbols), flat-gradient (green symbols) and red-cored (red symbols) galaxies as a function of $M_{\ast}$. From left to right, we show the results for the blue cloud, green valley and red-sequence. Errors are derived from 100 bootstrap resamplings.}
\end{figure*}

\section{Results}
\subsection{The color$-$magnitude diagram}

The color$-$magnitude diagram, or the color$-$mass diagram (CMD), is widespread used in the literature to diagnose the galaxy evolutional stage \citep{Faber 2007,Kaviraj 2007b,Schawinski 2014}. Based on their positions on the CMD, galaxies can be roughly categorized into two main populations, e.g., the red sequence galaxies and blue cloud galaxies.  Galaxies that lie between the "blue cloud" and "red sequence", are called the "green valley " (GV) galaxies \citep{Strateva 2001, Baldry 2004}. GV galaxies were traditionally thought to be the transition populations between the star-forming and the quenched galaxies \citep{Pan 2013, Mok 2013,Schawinski 2014, Pan 2014,Vulcani 2015}. To show the CMD clearly, we have divided the sample into 3 subsamples, according to their $\Delta$ NUV$-r$, where $\Delta$NUV$-r=$ (NUV$-r_{\rm r=3\arcsec .0}$)$-$(NUV$-r_{\rm total}$). NUV$-r_{\rm r=3\arcsec .0}$ and NUV$-r_{\rm total}$ are the color indices measured in the central $r=3\arcsec .0$ aperture and that of the whole galaxy, respectively. The typical error of $\Delta$ NUV$-r$, $\sigma_{0}$, is derived as $\int_{0}^{\sigma_{0}}N(\sigma)/N_{\rm gal}d\sigma$=0.80, where $N(\sigma)$ is the number of galaxies with a $\Delta$ NUV$-r$ error of $\sigma$. For this sample, $\sigma_{0}$ is 0.15 mag. In what follows, we called galaxies with $\Delta$ NUV$-r$>2$\sigma_{0}$ as "red-cored" galaxies. The "blue-cored" galaxies are those with $\Delta$ NUV$-r$<$-2\sigma_{0}$. They are denoted in red and blue symbols in panel a) of Figure. 2, respectively. Those galaxies with flat color gradients, e.g., galaxies with|$\Delta$ NUV$-r$|<2$\sigma_{0}$, are shown in green symbols. All AGNs in our sample are narrow-line (obscured) AGNs, so there should be no contribution from AGN continuum to the NUV$-r$ color index.
\begin{figure*}
\centering
\includegraphics[width=120mm,angle=0]{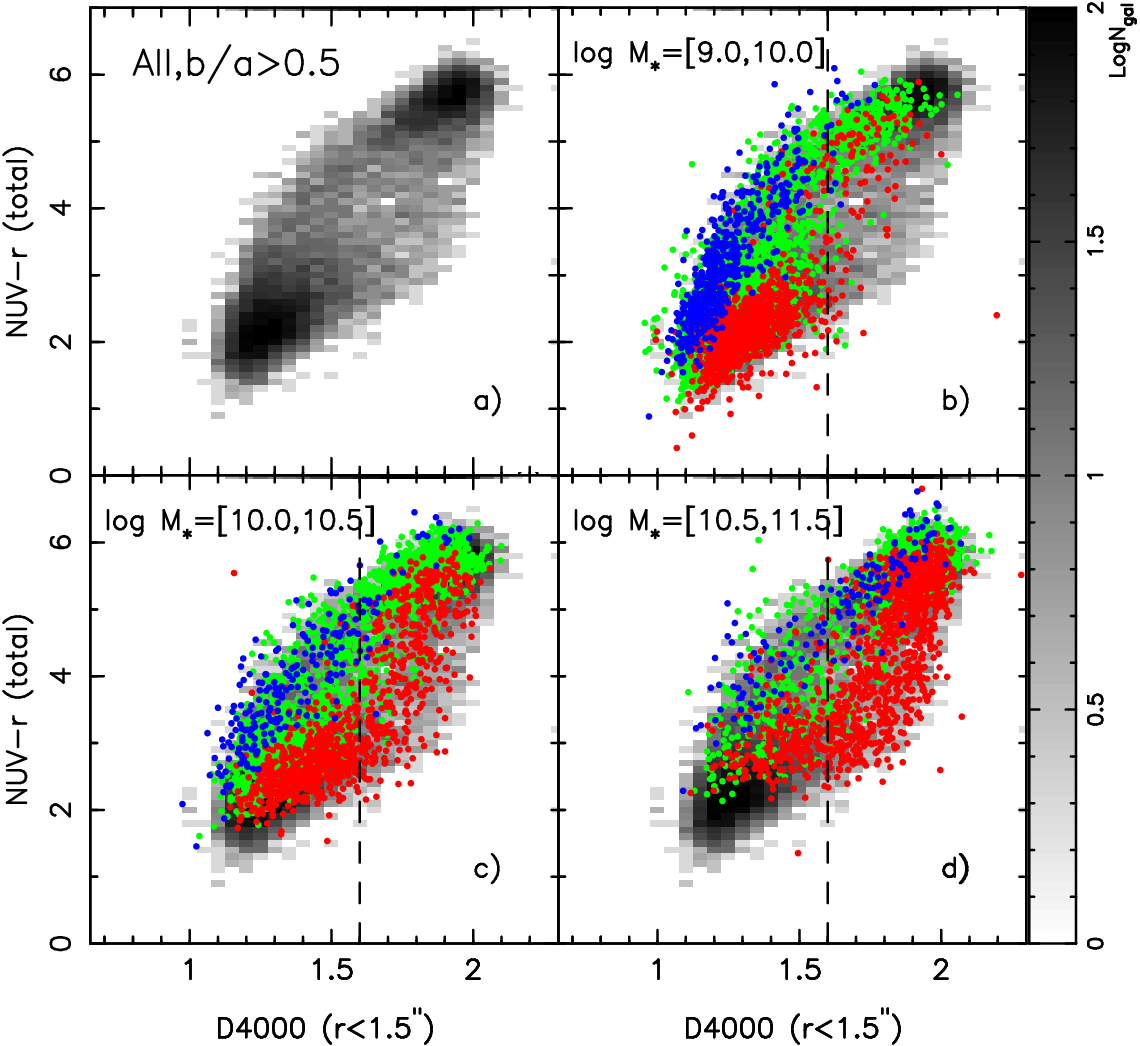}
\caption{panel a):The central $D_{n}4000$ .vs NUV$-r$ (total) relation for the whole sample.  panel b), panel c) and panel d) show this relation for the log $M_{\ast}/M_{\sun}$<10.0, log $M_{\ast}/M_{\sun}$=[10.0,10.5] and log $M_{\ast}/M_{\sun}$>10.5 subsamples, respectively. Color symbols are the same in Figure. 2.}
\end{figure*}

In panel b), panel c) and panel d) of Figure. 2, we show the $M_{\ast}-$ (NUV$-r$) diagram for the blue-cored, flat-gradient and red-cored galaxies, respectively. The gray scale represents the galaxy number density of the \emph{whole} sample. On the CMD, one can see the evident "color bimodality". The GV region is arbitrarily defined by the two pink lines, with $0.6\times M_{\ast}-2.4$<$(\rm NUV-r_{\rm})_{GV}$$<0.6\times M_{\ast}-1.4$, where $M_{\ast}$ is logarithm stellar mass. As shown in \cite{Kaviraj 2007a}, galaxies with an NUV$-r$ color less than 5.0 are very likely to have experienced recent star formation. In the blue cloud and GV region, the "blue-cored" phenomenon is thus corresponding to centrally concentrated star formation. In panel c), we show that flat-gradient galaxies distribute similarly with the whole sample on the CMD. This is expected since they are most numerous. Strikingly, panel d) shows that there are nearly no "red-cored" galaxies in the GV region below $M_{\ast}\sim 10^{10.0}M_{\sun}$.

In Figure.3, we show the fractions of the 3 kind galaxies, scaled with $M_{\ast}$, in the blue cloud, green valley and red sequence. It is clear from Figure. 3 that at $M_{\ast}>10^{10.5}M_{\sun}$, the blue cloud and GV region is mostly populated by red-cored galaxies. In the GV region, we see a significantly increase of blue-cored $M_{\ast}<10^{10.0}M_{\sun}$ galaxies when compared to the blue cloud, at the expense of a decrease of red-cored and flat-gradient galaxies. This can be interpreted as the outside-in quenching processes (regardless of its working mechanisms) transform a significant fraction of the red-cored and flat-gradient galaxies into blue-cored ones, and drive them evolve form blue cloud to GV. In the red sequence, galaxies are mostly with flat color gradients. This is expected since when NUV$-r$ color index is no longer sensitive to stellar age older than 1 $Gyr$ \citep{Kaviraj 2007b}. To summary, Figure. 2 and Figure. 3 confirms the findings of \citet{perez 2013} that $M_{\ast}<10^{10.0}M_{\sun}$ galaxies mainly grow "outside-in". The importance of the "inside-out" mode steadily increases with $M_{\ast}$, and becomes dominant at around $M_{\ast}>10^{10.5}M_{\sun}$.

\begin{figure*}
\centering
\includegraphics[width=120mm,angle=0]{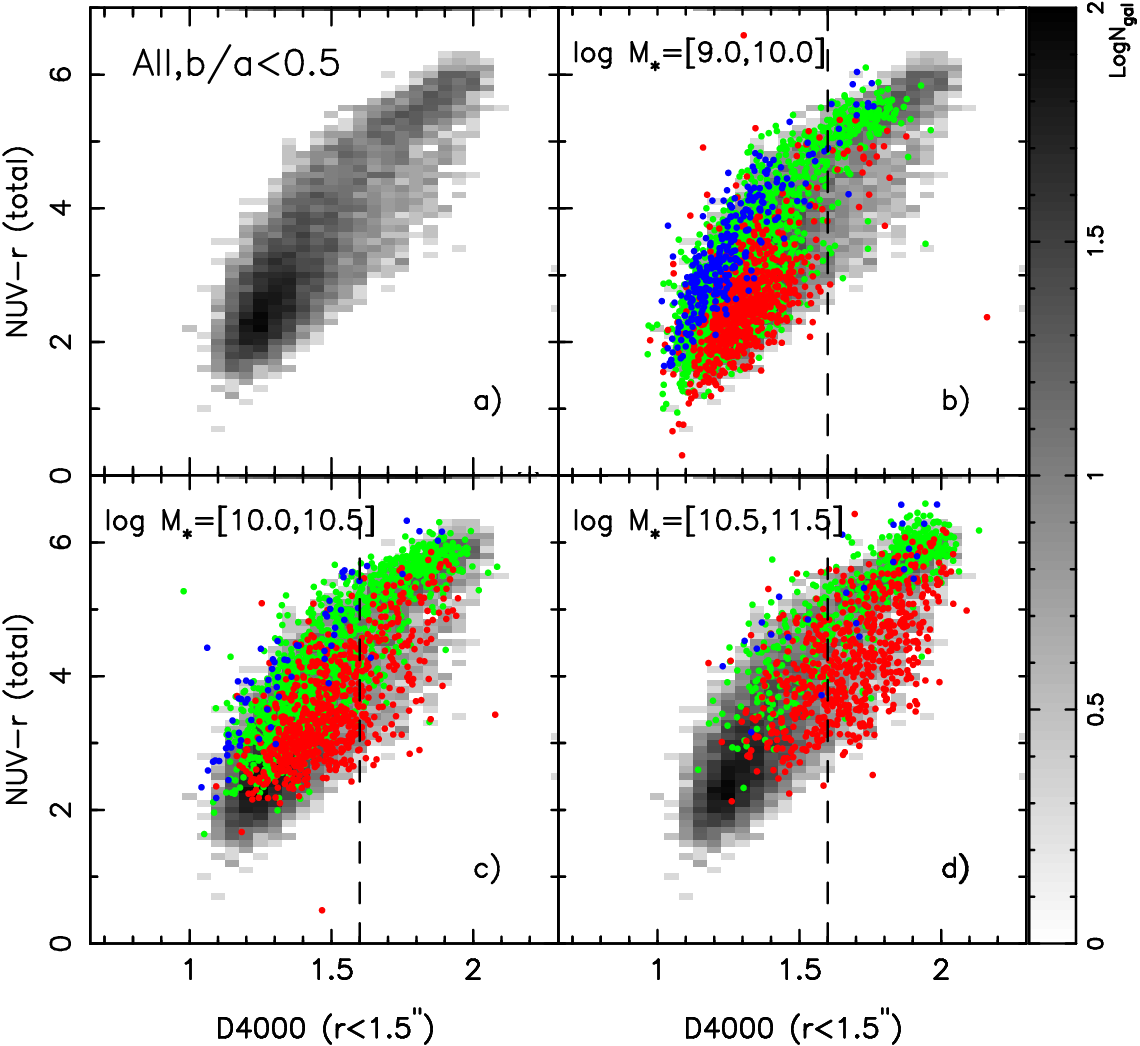}
\caption{Same to Figure. 4, but shown for the inclined galaxies.}
\end{figure*}

\subsection{The $D_{n}4000$ versus NUV$-r$ diagram}
In this part, we will use a spectroscopic index, the $D_{n}4000$ index measured on the SDSS $r=1\arcsec.5$ aperture spectra as an central stellar age indicator. The $D_{n}4000$ is widely used as a mean stellar age indicator in the literature, since it is sensitive to the ratio of present$-$ to past$-$ average star formation rate \citep{Kauffmann 2003a, Brinchmann 2004}, but is less affected by dust and metallicity.

In panel a) of Figure.4, we show the $D_{n}4000$ versus NUV$-r$ (total) relation for our sample. The grayscale represents the number density of galaxies. We can see that there is a positive relation between $D_{n}4000$ and NUV$-r$ (total), as also found in \citet{Kauffmann 2007}. The scatter of this relation reaches its maximum at NUV$-r\approx [3.0,5.0]$, .e.g. the GV region of the CMD. In panel b), panel c) and panel d), we show the the $D_{n}4000$ versus NUV$-r$ (total) relation for three subsamples, which are binned according to their $M_{\ast}$. The color coded symbols are the same as those in Figure. 2. In Figure. 4, one can find that the red and the blue symbols are still separable. This confirms that the NUV$-r_{\rm r=3\arcsec.0}$ is equivalent to the SDSS $D_{n}4000$ index.

In panel b) of Figure. 4, we find that, nearly all of the $M_{\ast}<10^{10.0}M_{\sun}$ galaxies have moved to the red sequence when their central $D_{n}4000$>1.6. In contrast, In panel d), we find that a large portion of the $M_{\ast}>10^{10.5}M_{\sun}$ galaxies still lie on the blue cloud or the GV region when their central $D_{n}4000$>1.6.  Panel b) and panel d) together show that, there seems to be two separable evolution paths through which the blue cloud galaxies evolve into the red sequence. In this work, it is naturally to interpret these two paths as the "outside-in" and the "inside-out" path, respectively.

In panel b), we show that the flat gradient galaxies are on the same path as those of "blue-cored" galaxies.  Thanks to the the accurate measurements of $D_{n}4000$ and NUV$-r$ (total), we can now conclude that at $M_{\ast}<10^{10.0}M_{\sun}$, the flat gradient galaxies follow a similar evolutionary path of the "blue-cored" when they move to the red sequence, e.g., they evolve "outside-in".

A potential issue which may affect the result of Figure. 4 is that we haven't applied dust reddening correction to the NUV$-r$ (total) color index. Since the color index of the inclined galaxies are more easily reddened by the galactic dust lanes, to assess the role of dust on our result, we show the $D_{n}4000$ versus NUV$-r$ (total) relation for the $b/a<0.5$ galaxies in Figure. 5. by comparing Figure. 4 and  Figure. 5, we find that the result is not changed for the $M_{\ast}< 10^{10.0}~M_{\sun}$ galaxies. At $D_{n}4000>1.6$, many $M_{\ast}> 10^{10.5}~M_{\sun}$ galaxies are reddened by dust, but their distributions can still be separated from those of the $M_{\ast}<10^{10.0}M_{\sun}$ on this diagram. This test suggests that the two separable paths revealed in Figure. 4 is intrinsic.

\section{Summary and discussion}
In this Letter, we investigate how galaxy mass assembly mode depends on stellar mass $M_{\ast}$, using a large sample of $\sim$10, 000 low redshift galaxies. The use of \emph{GALEX} photometry and SDSS data enables us to study a large sample without the IFS data. We compare the stellar age indicator measured in the galactic centers to that of the whole galaxy, to determine whether the galaxy grows "outside-in" or "inside-out".  We find that: in the $M_{\ast}-($ NUV$-r$) green valley, the $M_{\ast}<10^{10}~M_{\sun}$ galaxies are mostly "blue-cored" or have flat color gradients, whereas the $M_{\ast}>10^{10.5}~M_{\sun}$ galaxies are "red-cored". When their central $D_{n}4000$ index values exceed 1.6, the $M_{\ast}<10^{10.0}~M_{\sun}$ galaxies have lied on the UV red sequence, whereas a large fraction of the $M_{\ast}>10^{10.5}~M_{\sun}$ galaxies still lie on the UV blue cloud or the green valley region. These findings suggest that the main galaxy assembly mode is transiting from "the outside$-$in" mode to "the inside$-$out" mode at $M_{\ast}< 10^{10}~M_{\sun}$ and $M_{\ast}> 10^{10.5}~M_{\sun}$.

We interpret these results as a compromise between the internal and external process that driving star formation quenching in galaxies.  The internal process, like the AGN feedback \cite[e.g.,][]{Croton 2006} and the supernova feedback induced from the strong starburst activity \citep{Geach 2014}, can blow the gas out of the galactic center, causing the "inside-out" evolution. The external process, like the environment effects, will strip the gas content of galaxies firstly from their outskirts, causing the "outside-in" evolution. The capability that a galaxy can retrieve the lost gas, or accreting new gas that rebuilds its outskirts, is its gravity. It is thus not surprising to see the galaxy assembly modes have strong dependence on $M_{\ast}$.

Panel b) of Figure. 2 shows that a significant fraction of the $M_{\ast}< 10^{10.5}~M_{\sun}$ galaxies are "blue-cored". Previous works found that galaxies of this type are usually bulge-dominated systems, like the E+A galaxies \citep{Yang 2008} or the early type galaxies \citep{Lisker 2007, Suh 2010,Pan 2014}. This phenomenon can be either explained as the "survived" star formation of an late type galaxy which is suffering gas stripping, or explained as the "excess" central star formation, probably triggered in an merge event. Since the merge rate is rare in the local Universe, we suggest that the "survived star formation" explanation is the case for the majority of the "blue-cored" galaxies, especially for the low mass ones.

$M_{\ast}< 10^{10}~M_{\sun}$ galaxies are mostly satellites of massive halos. In a group/cluster environment, ram pressure stripping \citep{Gunn 1972} and tidal interactions, can remove the gas from galaxies, especially for those with low stellar mass densities \citep{Zhang 2013}. Other gentler processes such as "Starvation" and "Harassment", are also at work \citep{Barazza 2002,Weinmann 2009}. The important role of dense environment on the formation of red dwarf galaxies is confirmed by the works focused on the Virgo and Coma clusters (see \cite{Boselli 2014} for a review). In lower halo mass, \cite{Geha 2012} found that quenched $M_{\ast}< 10^{9}~M_{\sun}$ galaxies do not exist in the field, suggesting that the internal process alone is not sufficient in quenching star formation of low mass galaxies. At higher redshifts, using the COSMOS data, \cite{Pan 2013} found that the important role of environment on the evolution of $M_{\ast}< 10^{10}~M_{\sun}$ galaxies have been in place at $z<0.7$.

The hierarchical paradigm for galaxy formation implies that galaxies experience a variety of environments during their evolution, and may have complex SFH at their early assembly epoch. In Figure. 2 and Figure.4, we can find that there are many "red-cored" dwarf galaxies in the blue cloud. However, these galaxies are not quenched, even their cores are older than the global. It is worth to mention why we focus on the GV galaxies at this point. GV galaxies are more close to the "fully assembled" stage than the blue cloud galaxies (here we do not include those GV galaxies due to "rejuvenation"), hence holding important clues to the galaxy final assembly process. As can be seen in panel a) of Figure. 4, when galaxies evolved into the GV region, there seems to be two separable paths connecting the blue cloud and the red sequence. This is expected if there is a dichotomy in the galaxy assembly modes.

We find that the log $M_{\ast}/M_{\sun}$=[10.0,10.5] regime is a transition zone. This is not reported in \citet{perez 2013}, probably due to their small sample size. Since the sample used in this Letter is $\sim$ 100 times larger than that in \citet{perez 2013}, our results should be reliable in a statistical sense. At higher masses, our finding is in broad consistent with previous studies. The results of our work can be re-examined using the upcoming large data produced by the on-going IFS survey projects, such as CALIFA, MaNGA and SAMI in the near future.

\acknowledgments
We thank the anonymous referee for the constructive report. This work was supported by the NSFC projects (Grant Nos. 11473053, 11121062, 11233005, U1331201, 11225315, 1320101002, 11433005, and 11421303), the National Key Basic Research Program of China (Grant No. 2015CB857001), the ``Strategic Priority Research Program the
Emergence of Cosmological Structures'' of the Chinese Academy of Sciences (Grant No. XDB09000000), the Specialized Research Fund for the Doctoral Program of Higher Education (SRFDP, No. 20123402110037), and the Chinese National 973 Fundamental Science Programs (973 program) (2015CB857004).

\end{document}